\documentclass[10pt,aps,pra,twocolumn,showpacs,superscriptaddress,]{revtex4-1}

\usepackage{graphicx}
\usepackage{amsmath}
\usepackage{amssymb}
\usepackage{bm}
\usepackage{xcolor}
\usepackage[%
  colorlinks=true,
  urlcolor=blue,
  linkcolor=blue,
  citecolor=blue
]{hyperref}

\usepackage{siunitx}

\graphicspath{{./Figures_vsub/}}

\newcommand{\lkb}{Laboratoire Kastler Brossel, Coll\`ege de France, CNRS, ENS-PSL Research University, UPMC-Sorbonne Universit\'es, 11 place Marcelin Berthelot, 75005 Paris, France}

\begin{document}

\title{Clock spectroscopy of interacting bosons in deep optical lattices}

\author{R. Bouganne}
\altaffiliation{RB and MBA contributed equally to this work.} 
\affiliation{\lkb}

\author{M. Bosch Aguilera}
\altaffiliation{RB and MBA contributed equally to this work.} 
\affiliation{\lkb}

\author{A. Dareau}
\altaffiliation{Present address: Vienna Center for Quantum Science and Technology, TU Wien – Atominstitut, Stadionallee 2, 1020 Vienna, Austria}
\affiliation{\lkb}

\author{E. Soave}
\altaffiliation{Present address: Institut f\"ur Experimentalphysik, Universit\"at Innsbruck, 6020 Innsbruck, Austria}
\affiliation{\lkb}

\author{J. Beugnon}
\affiliation{\lkb}

\author{F. Gerbier}
\email{fabrice.gerbier@lkb.ens.fr}
\affiliation{\lkb}

\date{\today}

\begin{abstract}
We report on high-resolution optical spectroscopy of interacting bosonic $^{174}$Yb atoms in deep optical lattices with negligible tunneling. We prepare Mott insulator phases with singly- and doubly-occupied isolated sites and probe the atoms using an ultra-narrow ``clock'' transition. Atoms in singly-occupied sites undergo long-lived Rabi oscillations. Atoms in doubly-occupied sites are strongly affected by interatomic interactions, and we measure their inelastic decay rates and energy shifts. We deduce from these measurements all relevant collisional parameters involving both clock states, in particular the intra- and inter-state scattering lengths.
\end{abstract}

\maketitle

\section{Introduction}

In the last decade, progress in laser frequency stabilization and frequency comparison has led to a new generation of atomic clocks with unprecedented performances \cite{ludlow2015a}. These clocks use ultra-narrow $^1S_0\rightarrow {}^3P_0$ optical transitions (hereafter ``clock'' transitions), interrogated with ultra-stable lasers locked to high-finesse Fabry-Perot cavities. Because such optical transitions are essentially free of spontaneous emission, they provide new opportunities not only for frequency metrology, but also for quantum information processing \cite{gorshkov2009a,daley2011a}, many-body physics and quantum simulation \cite{gorshkov2010a,gerbier2010a,martin2013a,goldman2013b,rey2014a}. For instance, spin-orbit coupling has been demonstrated in $^{87}$Sr \cite{kolkowitz2017a} and $^{173}$Yb atoms \cite{livi2016a}. In these applications inter-atomic interactions play an important part. In particular, they are central to the study of quantum gases or the generation of entanglement. Recently, large spin-exchange interactions have been measured with fermionic $^{173}$Yb \cite{capellini2014a,scazza2014a}, which is promising for the simulation of quantum impurity models \cite{Riegger2017a}. While atomic interactions are usually detrimental in frequency metrology \cite{ludlow2015a}, degenerate quantum gases may help to improve the accuracy of optical clocks by offering a better control over interaction effects \cite{ludlow2015a}, as suggested by a recent demonstration with a degenerate Fermi gas of $^{87}$Sr \cite{campbell2017a}. Most optical clocks currently operate with fermionic atoms, which reduces but does not eliminate the problem of interactions at low temperatures \cite{campbell2009a}. However, bosonic atoms could provide advantages due to their reduced sensitivity to external fields and simpler level structure, or for comparing clocks based on different isotopes \cite{ludlow2015a}. Interaction effects could be essentially eliminated using many-body correlated states, \textit{e.g.} a Mott insulator with one atom per lattice site.

Narrow optical or microwave transitions can also been used to probe various properties of degenerate quantum gases. Bose-Einstein condensation in hydrogen was detected through the associated change in the optical spectrum recorded on the ${}^1S-{}^2S$ transition \cite{killian1998a}.
%The excitation spectrum and dynamic structure factors of bulk Bose-Einstein condensates were measured using two-photon Bragg spectroscopy.
This early study pointed out the role of intra- and inter-state interactions on the shape and position of the optical spectra. For bosonic gases in optical lattices, microwave spectroscopy has been used to probe the spatial structure of Mott insulator phases that arise for deep lattices \cite{foelling2006a,campbell2006a}. At zero temperature, the density profile shows extended regions of uniform integer filling, the so-called ``Mott plateaus'' \cite{bloch2008a}. Different fillings correspond to different interaction shifts that can be resolved spectroscopically. More recently, optical spectroscopy on the narrow ${}^{1}S_0\rightarrow {}^{3}P_2$ transition \cite{yamaguchi2010a} enabled to monitor the superfluid-Mott insulator transition in a gas of $^{174}$Yb atoms \cite{kato2016a}.

\begin{figure*}[ht!]
\centering
\includegraphics[width=\textwidth]{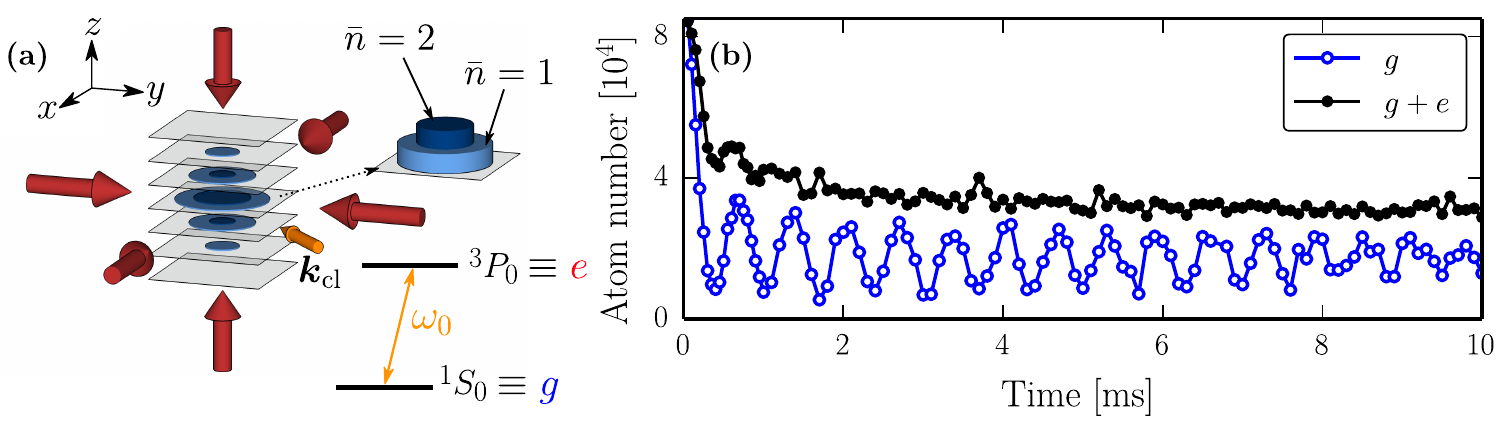}
\caption{\textbf{(a)} Sketch of the three-dimensional lattice geometry. The drawing also illustrates the density profile of the Mott insulator and the level scheme of the ultra-narrow clock transition connecting the ground state $g$ and the metastable excited state $e$. \textbf{(b)} Coherent driving on the clock transition in the deep Mott insulator regime. A coupling laser resonant on the clock transition is switched on at time $t=0$ with a Rabi frequency $\Omega/(2\pi)\approx 1500\,$Hz. Closed (respectively open) symbols represent the remaining total atom number (resp. the population in the ground state $g$).} 
\label{fig1}
\end{figure*}

In this work, we report on a spectroscopic study of a bosonic Mott insulator of $^{174}$Yb using the ${}^{1}S_0\rightarrow {}^{3}P_0$ ultra-narrow clock transition. Typical experimental results are shown in Fig. \ref{fig1}b. Starting from a sample in the ground state $g\equiv {}^{1}S_0$, a laser pulse resonant on the clock transition drives coherent oscillations between $g$ and the metastable excited state $e\equiv {}^{3}P_0$. After a fast initial decay of the total atom number, the contrast of the oscillations approaches unity and eventually decreases for $t \gtrsim 8\,$ms. This coherent dynamics can be understood by considering the spatial distribution of atoms in a trapped Mott insulator (sketched in Fig. \ref{fig1}a). For our experimental parameters, we expect a central core of doubly-occupied sites surrounded by an outer shell of singly-occupied sites. Inelastic collisions in the excited state lead to the fast initial decay, where all doubly-occupied sites are lost. When only singly-occupied sites remain, they display long-lived Rabi oscillations. In this article, we study in detail the impact of elastic and inelastic interactions on the coherent dynamics of the system. We show that this analysis can be used to measure the previously unknown collisional properties of $^{174}$Yb.

The paper is organized as follows. We describe the experimental setup in Sec. \ref{sec:experiment}, including the sequence used to prepare an ultracold bosonic gas in an optical lattice in Sec. \ref{sec:mottinsulator}, and the optical setup for the clock laser and detection scheme in Sec. \ref{sec:singleatomRabi}. We present a measurement of the on-site occupation numbers in the Mott insulator phase in Sec. \ref{sec:occupation}. In Sec. \ref{sec:interactions}, we analyze the dynamics of doubly-occupied sites. Using the model presented in Sec. \ref{sec:model}, we extract inelastic (Sec. \ref{sec:losses}) and elastic (Sec. \ref{sec:spectro}) interaction parameters from spectroscopic data. We conclude in Sec. \ref{sec:conclusion}.	

\section{Experimental setup and methods}
\label{sec:experiment}

\subsection{A Mott insulator of bosonic ytterbium atoms}
\label{sec:mottinsulator}
\subsubsection{Optical lattice setup}

Our experiment starts with a nearly pure Bose-Einstein condensate (BEC) of about $10^5$ $^{174}$Yb atoms in a crossed optical dipole trap (CDT) with initial trap frequencies $\{\omega_{x},\omega_{y},\omega_{z}\}=2\pi\times \{60\,\mathrm{Hz},230\,\mathrm{Hz},260\,\mathrm{Hz}\}$ (see \cite{dareau2015a,scholl2016a} for more details). The BEC is transferred into a three-dimensional cubic optical lattice depicted in Fig. \ref{fig1}a. This lattice results from the incoherent superposition of three orthogonal standing waves with orthogonal polarizations \cite{bloch2008a}. All lattice light beams derive from the same laser operating at the magic wavelength $\lambda_\mathrm{m}\approx 759.4\,$nm \cite{ludlow2015a,barber2008a}, for which the polarizabilities of $g$ and $e$ are equal \cite{dzuba2010a}. The periodic optical lattice potential experienced by the atoms irrespective of their internal state is then
\begin{eqnarray}
V_{\mathrm{OL}} & = \sum_{\alpha=x,y,z} V_{0,\alpha} \sin^2(k_\mathrm{m} \alpha),
\end{eqnarray}
where $V_{0,\alpha}$ are the lattice depths and where $k_\mathrm{m}=2\pi/\lambda_\mathrm{m}$ is the lattice wavenumber.
We calibrate the lattice depths $V_{0,\alpha}$ for given laser powers by Kapitza-Dirac diffraction. The vertical lattice (VL) depth $V_{0,z}\approx27\,E_\mathrm{R}$ is essentially fixed (see Sec. \ref{sec:loading}), while the horizontal lattice (HL) depths can be varied from zero to $ \{V_{0,x},V_{0,y}\}\approx\{24,25.4\}\,E_\mathrm{R}$. Here $E_\mathrm{R} = \hbar^2 k_\mathrm{m}^2/2M$ is the recoil energy, with $M$ the atomic mass. In this work, unless stated explicitly, we use the largest available lattice depths. The Gaussian envelope of the lattice lasers also entails an additional, slowly varying harmonic potential
\begin{eqnarray}
V_{\mathrm{T}} = \sum_{\alpha=x,y,z} \frac{1}{2}M\Omega_\alpha^2 \alpha^2.
\end{eqnarray}
By recording collective mode frequencies of a BEC, we infer the Gaussian beam waists for each standing wave, $\{w_x,w_y,w_z\} \approx \{$\SI{115}{\micro\meter},\SI{125}{\micro\meter},\SI{150}{\micro\meter}$\}$. The corresponding trap frequencies are $\{\Omega_x,\Omega_y,\Omega_z\}=2\pi\times \{42\,\mathrm{Hz},38\,\mathrm{Hz},33\,\mathrm{Hz}\}$ for lattice depths $\{V_{0,x},V_{0,y},V_{0,z}\}=\{24,25.4,27\}\,E_\mathrm{R}$.

\subsubsection{Adiabatic preparation of a Mott insulator}
\label{sec:loading}
The BEC is transferred into the optical lattice in three consecutive steps, aiming at preparing a stack of independent 2D Bose gases in the lowest Bloch band of the HL: (i) a fast ramp up of the VL, (ii) a slow extinction of the CDT and (iii) an adiabatic increase of the HL. 

In more detail, we first ramp up in 20\,ms a single standing wave propagating vertically (VL), superimposed on the CDT. This fast increase is used to freeze the vertical motion in the combined potential formed by the optical lattice and gravity. The potential $V_\mathrm{T}$ alone barely traps the atoms along gravity with a sag of the cloud much larger than in the CDT ($\Delta z_\mathrm{T} = g_0/\Omega_z^2 \approx$ \SI{230}{\micro\meter} compared to $\Delta z_\mathrm{CDT} = g_0/\omega_{z}^2 \approx $ \SI{4}{\micro\meter}, respectively, with $g_0$ the acceleration of gravity). The fast ramp up of the VL (as opposed to a slow, quasi-adiabatic transfer) prevents from inducing any motion along $z$ \cite{morsch2006a}, potentially leading to heating of the cloud. Note, however, that the duration of this first step is still long enough to avoid inter-band transitions. In a second step, the CDT is smoothly extinguished in 200\,ms to create a stack of independent two-dimensional condensates in the VL potential alone. This step reduces the atomic density and hence mitigates the rate of three-body losses (see Sec. \ref{sec:occupation}). It also avoids differential light shifts between the clock states created by the CDT lasers. In a final step, the two arms of the HL are increased to their desired values in 100\,ms, which roughly fulfills the criterion of \cite{gericke2007a} for adiabatic loading of the lowest Bloch band.

Each planar gas can be described by a 2D Bose-Hubbard Hamiltonian \cite{bloch2008a}
\begin{widetext}
\begin{eqnarray}\label{eq:BH}
\hat{H} &=& -\sum_{\langle i,j \rangle_\perp} J_\perp \hat{a}_{g,j}^\dagger \hat{a}_{g,i} 
 +\sum_{i_\perp}\frac{U_{gg}}{2}\hat{n}_i\left(\hat{n}_i-1\right)+\sum_{i_\perp}\frac{M}{2}(\Omega_x^2 x_i^2+\Omega_y^2 y_i^2) \hat{n}_i.
\end{eqnarray}
\end{widetext}
Here the notation $i_\perp$ indicates summation over all lattice sites at positions $\bm{\rho}_i= (x_i,y_i)$ in the $x-y$ plane, and $\langle i,j \rangle_\perp$ in-plane tunneling to nearest-neighbors with matrix element $J_\perp$. Tunneling along the gravity direction has been neglected. The on-site energy $U_{gg}$ is given by
\begin{eqnarray}\label{eq:Ugg}
U_{gg} = \frac{4\pi \hbar^2 a_{gg}}{M} \int \mathrm{d}z\,\mathrm{d}\bm{\rho}\,
\Big\vert W_z(z) W_\perp(\bm{\rho}-\bm{\rho}_i)\Big\vert^4,
\end{eqnarray}
where $\bm{\rho}$ denotes a two-dimensional vector in the $x-y$ plane and where $W_z$ (respectively $W_\perp$) are the Wannier functions for the vertical (resp. horizontal) lattice potential. The scattering length $a_{gg}$ describes low-energy \emph{s}-wave scattering between two atoms in the internal state $g$, and has been measured in \cite{kitagawa2008a}, $a_{gg} = 105(2)\,a_0$ with $a_0$ the Bohr radius. Using collapse and revival dynamics (see \cite{greiner2002b} and appendix), we measure $U_{gg}/h = 1475(25)\,$Hz, which compares well with the value $1420\,$Hz calculated from Equ. \ref{eq:Ugg} using the calibrated lattice depths. Unless otherwise mentioned, the quoted error bars represent a $68\,\%$ confidence interval on the optimum fit value.

At zero temperature, the 2D Bose-Hubbard model supports phase transitions to incompressible Mott insulator (MI) phases \cite{bloch2008a,spielman2007a,capogrosso2008a,jimenezgarcia2010a}. For a filling factor $\overline{n}=1$, \emph{i.e.} 1 atom per lattice site, Monte-Carlo simulations predict a transition for a critical value $(U_{gg}/J_\perp)_c \approx 16.7$ \cite{capogrosso2008a}, corresponding to HL depths around $V_{0,x}\approx V_{0,y}\approx 9\,E_\mathrm{R}$. The smooth harmonic trap $V_\mathrm{T}$ leads to a characteristic ``wedding cake'' density profile, \textit{i.e.} density plateaus with integer filling, the denser plateaus occurring near the trap center \cite{bloch2008a}. In the present work, the HL depths become large enough at the end of the loading procedure so that we can safely neglect tunneling altogether ($J_\perp \approx 0$).
Introducing a chemical potential $\mu$, the spatial structure is then given, in the local density approximation, by
\begin{eqnarray}\label{eq:nBH}
n(\bm{\rho}_i) = \mathrm{Int}\left[\frac{\mu- \frac{M}{2}(\Omega_x^2 x_i^2+\Omega_y^2 y_i^2)}{U_{gg}}\right]+1,
\end{eqnarray}
where $\mathrm{Int}(x)$ denotes the integer part of $x$. For finite temperatures and/or tunneling (still small compared to $U_{gg}$), the overall density profile remains similar but with smoother edges than predicted by Equ. \ref{eq:nBH}. The relative weight of the plateau with $\overline{n}$ atoms can be characterized by its population normalized to the total atom number, noted $\mathcal{F}_{\overline{n}}$ in the remainder of the article.

\subsection{Single-atom Rabi oscillations on the clock transition}
\label{sec:singleatomRabi}
With a MI in the ground state $g$ as a starting point, we now describe our experiments involving spectroscopy on the clock transition. In this section, we focus on the simplest case with only one atom per site. This can be realized by loading a sufficiently small number of atoms (such that only a plateau with $\overline{n}=1$ atom per site appears) as in Fig. \ref{fig2}a.
%, or by concentrating on the long time dynamics of Rabi oscillations as in Fig. \ref{fig1}b. 

\subsubsection{Optical setup}
The optical setup has been presented in detail in \cite{dareau2015a}. Briefly, a laser resonant with the $g-e$ clock transition near $\lambda_{0}\approx 578.4\,$nm is locked on a high-finesse cavity well-isolated from its surroundings which serves as frequency reference. We split the optical path into one going to the cavity using an optical fiber with active Doppler noise cancellation \cite{ma1994a}, and the other going towards the atomic cloud. The wavevector $\bm{k}_\mathrm{cl}$ is oriented along the $\bm{e}_x+\bm{e}_y$ direction (see Fig. \ref{fig1}a).

The $g-e$ electric dipole transition is forbidden at zero field ($J=0 \rightarrow J'=0$ transition, with $J$ the total electronic angular momentum). Following the method pioneered in \cite{taichenachev2006a,barber2006a}, we use a static magnetic field $\bm{B}=B_0 \bm{e}_z$ (with $B_0 \approx 182\,$G) to enable an effective electric dipole coupling between $g$ and $e$. Neglecting motional degrees of freedom, the effective coupling strength is $\Omega_\mathrm{cl} \propto B_0 \sqrt{I_\mathrm{cl}}$ \cite{taichenachev2006a}, where $I_\mathrm{cl}=2P_\mathrm{cl}/(\pi w_\mathrm{cl}^2)$, $P_\mathrm{cl}$ and $w_\mathrm{cl}$ are respectively the intensity, power and waist ($1/\mathrm{e}^2$ radius) of the clock laser beam. For atoms in a deep optical lattice, the complete transition amplitude (hereafter denoted as ``Rabi frequency'' for simplicity) is $\Omega=\kappa\Omega_\mathrm{cl}$, where the additional factor $\kappa= \vert \langle W_\perp \vert \mathrm{e}^{\mathrm{i} \bm{k}_\mathrm{cl} \cdot \bm{\hat{r}}} \vert W_\perp \rangle\vert  \approx 0.9$ takes into account the overlap between motional states \cite{ludlow2015a}. We find $\Omega\approx 2\pi\times 1500\,$Hz using $I_\mathrm{cl}\approx2.4\times 10^5\,$mW/cm$^2$ ($P_\mathrm{cl}\approx 18.5\,$mW and $w_\mathrm{cl}\approx$ \SI{70}{\micro\meter}) and a laser linearly polarized along $\bm{e}_z$, in good agreement with the measured value $\Omega \approx 2\pi \times 1500\,$Hz for the experiment of Fig. \ref{fig1}b. Hopping transitions to different sites and to higher bands can be safely neglected (the Lamb-Dicke parameter is small, $(k_\mathrm{cl} a_{\mathrm{lat}})^{2}\approx 0.08$ with $a_{\mathrm{lat}}\approx 26\,$nm the typical extent of the atomic wavefunction $W_\perp$).

\subsubsection{Detection}
\label{sec:detection}
We detect atoms in the ground state $g$ using standard resonant absorption imaging on the $^1S_0\rightarrow {}^1 P_1$ transition. Atoms in the excited state $e$ are detected using a repumping laser on the $^3P_0\rightarrow {}^3D_1$ transition near $1389\,$nm. Here and in the remainder we denote by $N_{g/e}$ the atom number in the state $g/e$. The auxiliary ${}^3D_1$ state can decay to the $^3P_J$ states ($J=0,1,2$), where $^3P_{0,2}$ are metastable and where $^3P_1$ decays to the ground state by emitting a photon at $556\,$nm.
The metastable $^3P_2$ state is a dark state for both the repumping and the imaging lasers. However, the branching ratio $^3D_1 \rightarrow {}^3P_2$ is small \cite{bowers1996a}, so that the repumping efficiency $\eta_e$ from $^3P_0$ to $^1S_0$ is close to unity for a \SI{500}{\micro\second}-long repumping pulse. Comparing $N_g$ and $N_e$ for Rabi oscillations similar to Fig. \ref{fig2}a, we estimate $\eta_e \approx 80\,$\%, consistent with a calculation based on optical Bloch equations. In addition, atoms in the ground state are very far off-resonant and hardly affected. Hence, applying the repumping pulse allows us to detect the total population $N_g+N_e$. To detect selectively atoms in $e$ and measure only $N_e$, we apply an additional \SI{5}{\micro\second}-long removal pulse on the $^1S_0\rightarrow {}^1 P_1$ transition before the repumping pulse. Atoms in $g$ scatter many photons and are pushed away from the imaging region. Atoms in $e$ are mostly unaffected by the removal pulse, although we measure a reduced detection efficiency by approximately $10\,$\%. This is possibly due to secondary scattering between trapped $e$ atoms and untrapped $g$ atoms leaving the sample.
We have also observed a slight influence of the atomic density on the repumping efficiency. The observed efficiency is reduced by roughly $10\,$\% when $N_e$ is higher than about $3\times10^4$. We do not correct experimental atom numbers in $e$ for the repumping efficiency but we take it into account in the analysis of Sec. \ref{sec:losses} and Sec. \ref{sec:spectro}.

\subsubsection{Single-atom dynamics}
\label{sec:singleatom}

For atoms in singly-occupied lattice sites, the time evolution consists of textbook Rabi oscillations between $g$ and $e$, described by the Hamiltonian
\begin{eqnarray}
\label{eq:Hn1}
\hat{H}_\mathrm{eff}^{(\overline{n}=1)}&=
\left[\begin{array}{cc}
0 & \displaystyle \frac{\hbar\Omega}{2} \\
\displaystyle \frac{\hbar\Omega}{2} & -\hbar\delta 
\end{array}\right],
\end{eqnarray}
with $\delta=\omega_\mathrm{cl}-\omega_0$ the laser detuning, $\omega_0=2\pi c/\lambda_0$ the atomic Bohr frequency of the transition, $\omega_\mathrm{cl}$ the clock laser frequency and $\Omega \in \mathbb{R}$ the Rabi frequency introduced before. Starting from a sample in the ground state and switching on the coupling laser, the atomic populations oscillate between $g$ and $e$ at the frequency $\sqrt{\Omega^2+\delta^2}$. In the remainder we characterize the coupling laser pulse by its duration $T$, or equivalently by its \textit{area} $\Omega T$. We show in Fig. \ref{fig2}a Rabi oscillations for a cloud prepared with only singly-occupied sites [$N_\mathrm{at}\approx 8\times 10^3$], where almost full contrast is observed up to 10\,ms. However, for higher atom numbers as in Fig. \ref{fig1}b [$N_\mathrm{at}\approx 8\times 10^4$], Rabi oscillations are damped on the same time scale. For both cases the Rabi frequency is $\Omega/(2\pi)\approx 1500\,$Hz. In the following we consider possible dephasing mechanisms to explain this difference.

We first consider a possible deviation from the exact magic wavelength, inducing a position-dependent differential light shift caused by the lattice potential. A calculation based on \cite{barber2008a} indicates a differential shift across the cloud smaller than $10\,$Hz for a wavelength mismatch of $0.1\,$nm. Hence, this effect can be neglected in the present work.

Frequency or intensity fluctuations of the clock laser also induce dephasing over time. Although the intensity of this laser is not actively stabilized at the atomic cloud location, careful monitoring shows that intensity fluctuations remain below 1\% and cannot explain the observed dephasing. Regarding frequency fluctuations, we recorded spectra at low atom numbers (to ensure almost unity filling across the cloud), $\pi$-pulse areas and long pulse times. When the pulse time exceeds $\sim 10\,$ms, we observe substantial shot-to-shot fluctuations of the measured transition probability with identical parameters. This could be explained by shot-to-shot fluctuations of the clock laser frequency with a standard deviation of about $100\,$Hz, presumably due to the high-finesse cavity. This value can certainly be improved in future work. In order to evaluate the impact of such frequency fluctuations on the coherent dynamics in Fig. \ref{fig1}b, we modeled them by a random detuning $\delta$ with a Gaussian distribution function of width $\Delta\omega_\mathrm{cl}$. We verified numerically that a width $\Delta\omega_\mathrm{cl}$ greater than $2\pi\times 600$\,Hz would be required to account entirely for the observed damping, a value incompatible with the narrowest measured spectra.

Additional dephasing mechanisms come from the Gaussian profile of the coupling beam. This profile entails a non-uniform $\Omega$ and also induces a position-dependent differential light shift, leading to a non-uniform $\delta$ \cite{barber2006a}. Close to resonance, the first effect is dominant. Inhomogeneities cause a dephasing over time between the center and the edges of the cloud, and thus to an apparent damping when averaging over the whole cloud. The dephasing time $\tau_\mathrm{d}$ thus depends on the cloud size. Using a parabolic approximation for the Gaussian profile, we estimate $\tau_\mathrm{d} \approx \alpha w_\mathrm{cl}^2/(\Omega R^2)$, where $R$ is the radius of the cloud and $\alpha$ a numerical factor depending on the atomic density distribution\footnote{Although a misalignment of the clock beam could also increase the damping rate, this would entail a reduction of the maximum achievable Rabi frequency, which we do not observe.}. We used the model for the spatial atomic density of Sec. \ref{sec:occupation} to obtain the cloud size for a given $N_\mathrm{at}$ and to estimate $\alpha\approx6$.  For the highest atom number $N_\mathrm{at}\approx 8\times 10^4$ [$R\approx $ \SI{19}{\micro\meter}] we find $\tau_\mathrm{d} \approx 9\,$ms, close to the observed damping time in Fig. \ref{fig1}b. For the lowest atom number $N_\mathrm{at}\approx 8\times 10^3$ [$R\approx $ \SI{7.6}{\micro\meter}] we find $\tau_\mathrm{d} \approx 50\,$ms, consistent with almost no decay observed in Fig. \ref{fig2}a.

We conclude that inhomogeneous dephasing is likely responsible for the observed damping for large $N_\mathrm{at}$ but quickly becomes negligible when $N_\mathrm{at}$ decreases due to the quadratic dependence $\tau_d \sim R^{-2}$. In the low $N_\mathrm{at}$ limit, the damping is then most likely dominated by frequency fluctuations on a time scale $\gtrsim\,10$ms. We note that the damping has little influence on the shape and position of the spectra shown in Sec. \ref{sec:spectro}, where the pulse time $T$ is shorter than the damping time. However, the single-particle damping discussed in this paragraph limits the uncertainty on our measurement of collisional parameters presented in Sec. \ref{sec:spectro}.

\subsection{Determination of Mott shell populations}
\label{sec:occupation}
\begin{figure*}[hpt]
\centering
\includegraphics[width=\textwidth]{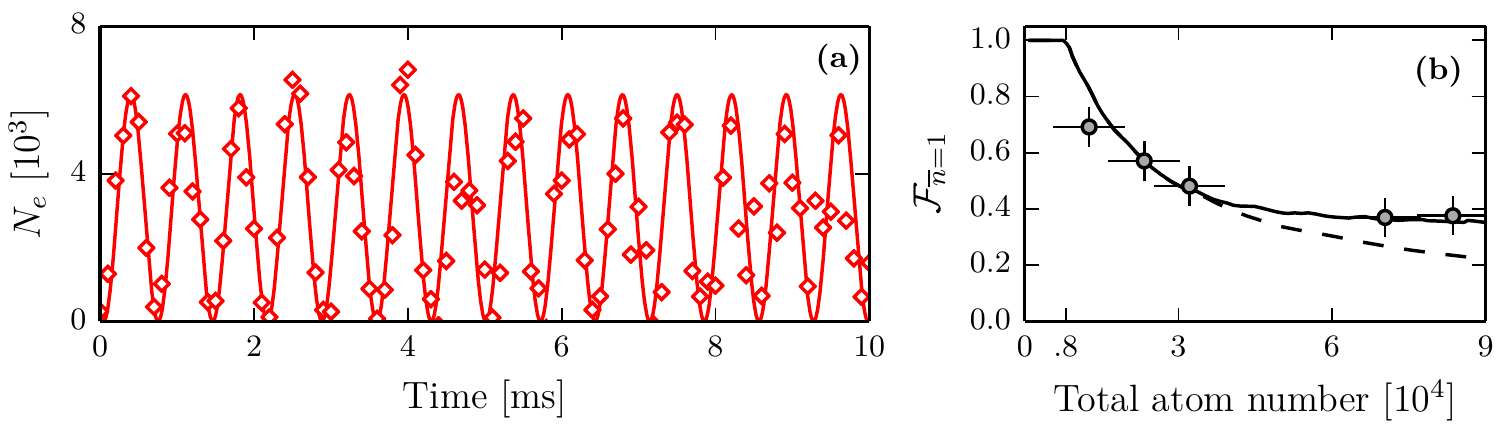}
\caption{\textbf{(a)} Coherent driving on the clock transition with Rabi frequency $\Omega/(2\pi)\approx 1500\,$Hz, recording the population in $e$. The atom number $N_\mathrm{at}\approx 8\times 10^3$ corresponds to a single Mott shell with unity filling. We observe long-lived oscillations up to 10\,ms. The solid line is a sinusoidal fit to the data. \textbf{(b)} Normalized population $\mathcal{F}_{\overline{n}=1}$ of the Mott shell with single occupancy as a function of the total atom number $N_\mathrm{at}$. Data points are extracted from the asymptotic behavior of coherent oscillations as in Fig. \ref{fig1}b. The dashed line is the prediction from our loading model assuming adiabaticity and zero temperature (see Sec. \ref{sec:occupation}). The solid line also includes three-body collisions that quickly empty sites with triple occupancy.} 
\label{fig2}
\end{figure*}	

Fig. \ref{fig1}b illustrates that the temporal dynamics at early times can strongly differ from the Rabi oscillations expected for singly-occupied sites alone, as in Fig. \ref{fig2}a. While elastic interactions provide state-dependent energy shifts, inelastic interactions lead to a fast decay leaving only singly-occupied sites after $\sim 1\,$ms (see Sec. \ref{sec:losses}). The long-time asymptote of $N_e+N_g$ in Fig. \ref{fig1}b thus reflects the initial fraction $\mathcal{F}_{\overline{n}=1}$ of atoms in singly-occupied sites of the MI phase. In Fig. \ref{fig2}b, we show the measured values of this asymptote for various initial atom numbers $N_\mathrm{at}$.

In order to compare this measurement to the expected value of $\mathcal{F}_{\overline{n}=1}$ in a deep MI phase, we model the first phase of our loading sequence where the VL is quickly increased (see Sec. \ref{sec:loading}) using a sudden approximation. The initial atomic distribution for a BEC in the CDT is projected on a periodic potential of period $d = \lambda_\mathrm{m}/2$ much smaller than the initial Thomas-Fermi half-length $L$ of the BEC. For a BEC in the Thomas-Fermi regime \cite{dalfovo1999a}, this results in a distribution $N(z_k) \approx (1-z_k^2)^2(15 N_\mathrm{at}d)/(16L)\theta\left[1-z_k^2\right]$ for the number of atoms in each plane of the VL at the altitude $z_k=kd/L$ (the integer $k$ labels the plane), with $\theta$ the Heaviside function. Using this distribution, we then assume a zero entropy sample in each plane, determined as the $J_\perp=0$ ground state of the 2D Bose-Hubbard model in Equ. \ref{eq:BH} with $N(z_k)$ atoms. We finally compute the normalized populations summing over all $k$. 

For the parameters of our experiment, there are typically $2 L/d \approx 10-12$ occupied planes, with occupation numbers in a deep MI phase ranging from $\overline{n}=1$ to $\overline{n}=3$. For the lowest atom numbers explored in this work [$N_\mathrm{at}\approx 8\times 10^3$], we find only a plateau with $\overline{n}=1$, in agreement with Fig. \ref{fig2}a. For the largest atom numbers [$N_\mathrm{at}\approx 8\times 10^4$], we find plateaus with normalized populations $\mathcal{F}_{\overline{n}=\{1,2,3\}} = \{0.25, 0.39, 0.36\}$. The prediction of this loading model for the normalized population $\mathcal{F}_{\overline{n}=1}$, shown in Fig. \ref{fig2}b as dashed line, agrees with the measured values only for low atom numbers $N_\mathrm{at}\lesssim 4\times 10^4$. 

We attribute the marked difference for higher atom numbers to three-body inelastic losses, that occur at relatively high rates in optical lattices. We estimate a lifetime $\tau_\mathrm{3B} \approx 100\,$ms for triply-occupied sites using the three-body rate constant $L_3 \approx 7\times10^{-30}\,$cm$^6$/s measured in \cite{fukuhara2009a}, comparable to the loading time in the HL. A detailed kinetic modeling of these losses during the loading sequence is beyond the scope of this work. Here we extend our model in the simplest possible way, by assuming that all triply-occupied sites have decayed during the loading and are therefore empty when the measurements are performed (the model predicts negligible population of sites with occupancy $\overline{n}>3$, and we neglect them in our discussion). For the largest atom numbers [$N_\mathrm{at}\approx 8\times 10^4$], the normalized populations become $\mathcal{F}_{\overline{n}=\{1,2,3\}} = \{0.36, 0.64, 0\}$. The prediction of this lossy loading model, shown in Fig. \ref{fig2}b as solid line, agrees well with the measured values and suggests that the in-trap density distribution is close to the predicted one.

\section{Interacting atoms driven on the clock transition}
\label{sec:interactions}

\subsection{Model}
\label{sec:model}
We now consider the dynamics of doubly-occupied sites driven by the coupling laser, which differs from singly-occupied sites in several aspects. First, due to bosonic enhancement, the coupling strength is $\sqrt{2}$ times higher for double than for single occupancy. Second, the three possible symmetric states $\vert gg \rangle$, $\vert eg \rangle$ and $\vert ee \rangle$ have in general different interaction energies, characterized by Hubbard parameters $U_{gg}$, $U_{eg}$ and $U_{ee}$, the last two being unknown. Finally, the states $\vert eg \rangle$ and $\vert ee \rangle$ are prone to inelastic decay via principal quantum number changing collisions. We model this inelastic process by adding an imaginary term $-\mathrm{i} \hbar \gamma_{e\alpha}/2$ to the Hamiltonian, with $\alpha=e,g$. This results in a dynamics captured by a non-Hermitian effective Hamiltonian
\begin{widetext}
\begin{eqnarray}
\label{eq:Hn2}
\hat{H}_\mathrm{eff}^{(\overline{n}=2)}&=
\left[\begin{array}{ccc}
0 & \displaystyle \frac{\hbar\Omega}{\sqrt{2}} & 0\\
\displaystyle \frac{\hbar\Omega}{\sqrt{2}} & \displaystyle U_{eg}-U_{gg}-\mathrm{i}\frac{\hbar\gamma_{eg}}{2}-\hbar\delta  & \displaystyle \frac{\hbar\Omega}{\sqrt{2}}\\
0 & \displaystyle \frac{\hbar\Omega}{\sqrt{2}} & \displaystyle U_{ee}-U_{gg}-\mathrm{i}\frac{\hbar\gamma_{ee}}{2}-2\hbar\delta
\end{array}\right]
\end{eqnarray}
in the $\{\vert gg \rangle, \vert eg \rangle,\vert ee \rangle\}$ basis. 
\end{widetext}

We numerically solve the generalized Schr\"odinger equation using the effective Hamiltonian Equ. \ref{eq:Hn2} with initial condition $\vert \Psi^{(2)} \rangle = \vert gg \rangle$. We also solve the Schr\"odinger equation for singly-occupied sites using Equ. \ref{eq:Hn1} with initial condition $\vert \Psi^{(1)} \rangle = \vert g \rangle$. This gives transition probabilities denoted $P_{\alpha}^{(2)}=\vert \langle \alpha \vert \Psi^{(2)} \rangle \vert^2$ with $\alpha=gg,eg,ee$ and $P_{\beta}^{(1)}=\vert \langle \beta \vert \Psi^{(1)} \rangle \vert^2$ with $\beta=g,e$. We then sum the contributions of doubly- and singly-occupied sites to obtain the average populations $\overline{N}_g$ and $\overline{N}_e$. For example, we have
\begin{eqnarray}
\label{eq:Ne}
\frac{\overline{N}_e}{N_\mathrm{at}} & = \eta_e\mathcal{F}_{\overline{n}=1} P_{e}^{(1)} + \eta_e \mathcal{F}_{\overline{n}=2} \left( P_{ee}^{(2)} + \frac{ 1}{2} P_{eg}^{(2)}\right).
\end{eqnarray}
We assume the repumping efficiency $\eta_e$ to be independent from the filling factor for simplicity.

\subsection{Lifetime of doubly-occupied sites}
\label{sec:losses}

\begin{figure*}[ht!]
\centering
\includegraphics[width=\textwidth]{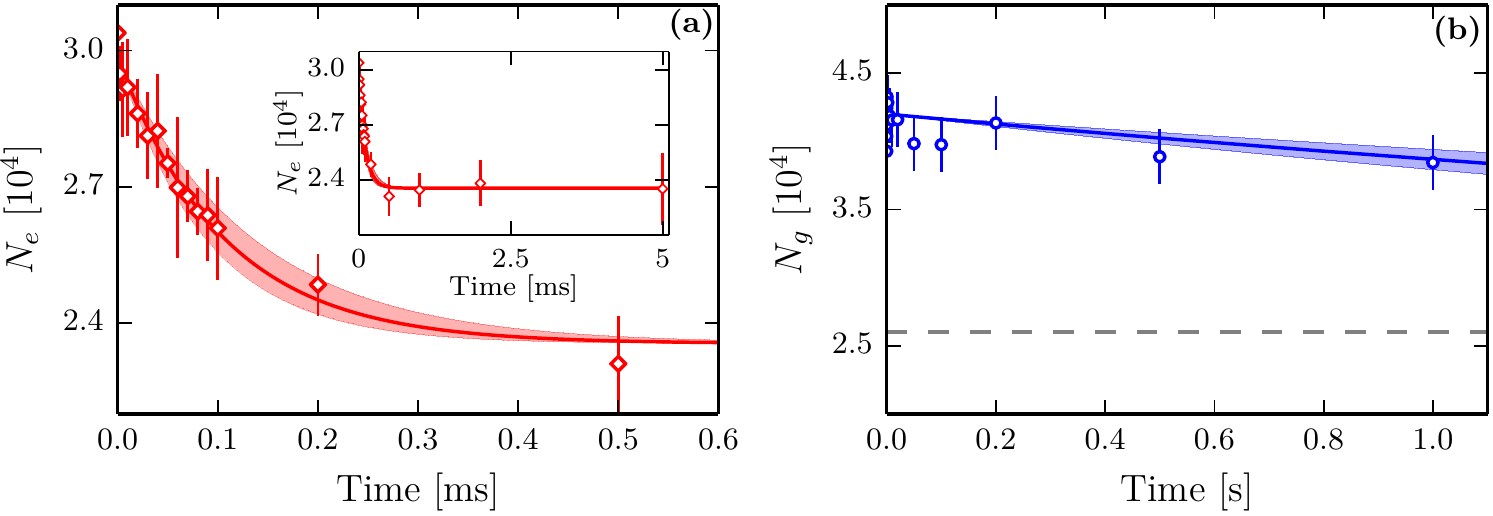}
\caption{\textbf{(a)} Lifetime measurement for a sample with only atoms in $e$. Doubly-occupied sites quickly decay through inelastic collisions. The inset shows the plateau of remaining singly-occupied sites for longer times. \textbf{(b)} Absence of inelastic collisions involving $g$ and $e$. The dashed line shows the asymptote expected for a complete decay of $e-g$ pairs. In \textbf{(a)} and \textbf{(b)}, solid lines are exponential fits to the data, with the shaded area reflecting the $68\,$\% confidence intervals.} 
\label{fig3}
\end{figure*}

In this section we measure the inelastic loss rates $\gamma_{ee}$ and $\gamma_{eg}$. We first investigate the role of $e-e$ inelastic collisions. After a coupling laser pulse of area $\Omega T\approx\pi$ [$\Omega/(2\pi)\approx 1500$\,Hz] in order to obtain a substantial population of $\vert ee \rangle$, we apply a removal pulse (see Sec. \ref{sec:detection}) to get rid of remaining atoms in state $g$. We are then left with a collection of singly- and doubly-occupied sites where all atoms are in the excited state $e$. We show in Fig. \ref{fig3}a the measured lifetime of this sample. We detect a fast exponential decay at short times which we interpret as the consequence of inelastic $e-e$ collisions. For longer times, we observe a plateau corresponding to the remaining $e$ atoms in singly-occupied sites. The exponential decay rate is a direct measurement of $\gamma_{ee} = 9300(100) \,\mathrm{s}^{-1}$.

A similar technique is used to investigate the role of $e-g$ inelastic collisions. We apply a coupling laser pulse of area $\Omega T\approx\pi/2$ [$\Omega/(2\pi)\approx 1500$\,Hz] in order to obtain a substantial population of $\vert eg \rangle$. We measure the atom number in the ground state $g$ which is expected to decay as $\dot{\overline{N}}_g/N_\mathrm{at} = - \gamma_{eg} \mathcal{F}_{\overline{n}=2} P_{eg}^{(2)}/2$. Fig. \ref{fig3}b shows a typical measurement, where almost no losses occur even after one second. In order to extract a damping rate, we fix the initial populations using the model in Sec. \ref{sec:model}. An exponential fit to the data, with a rate $\gamma$ as the only free parameter, yields $\gamma = 0.5(1) \,\mathrm{s}^{-1}$. The projected asymptote of the decay is represented with the dashed line in Fig. \ref{fig3}b. We measure a similar lifetime for atoms in $g$ in the absence of the coupling laser. Hence the measured damping rate $\gamma$ is only an upper bound for $\gamma_{eg}$.

\subsection{Spectroscopy of elastic interactions}
\label{sec:spectro}

\begin{figure*}[!t]
\centering
\includegraphics[width=\textwidth]{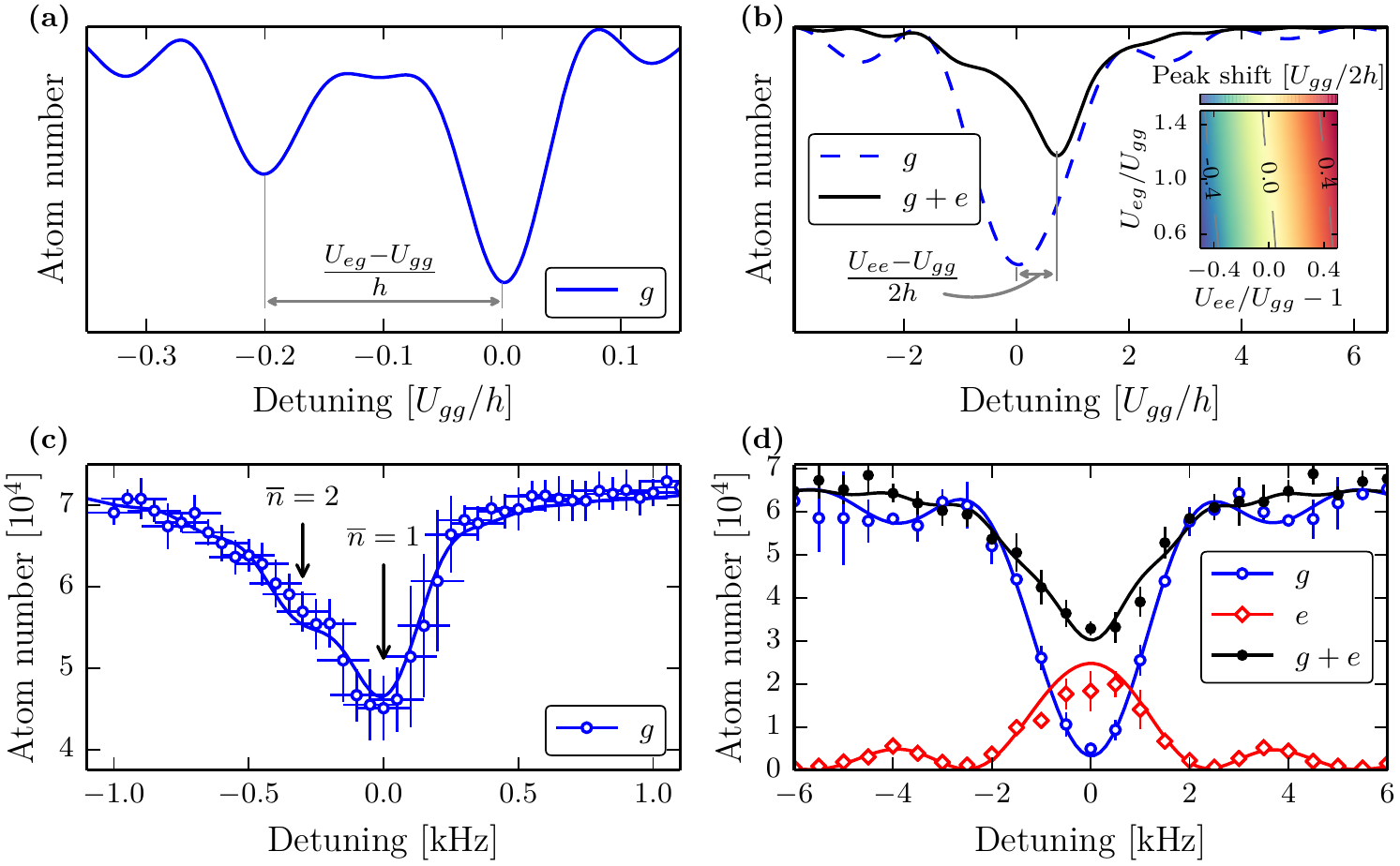}
\caption{
Determination of $U_{eg}$ and $U_{ee}$, (a-b) illustrate the methods, (c-d) show the measurements.
\textbf{(a)} Method for measuring $U_{eg}$. The number of atoms in $g$ is plotted with respect to detuning, with a pulse area $\Omega T=\pi$. Singly-occupied sites are excited on the single-atom resonance near $\delta=0$, with $\delta$ the laser detuning. Interactions shift the resonance for doubly-occupied sites to $\delta=(U_{eg}-U_{gg})/\hbar$. This interaction sideband can be resolved with sufficiently weak Rabi frequency $\Omega \ll \vert U_{eg}-U_{gg} \vert/\hbar$. For illustrative purposes, $\Omega/(2\pi) = 70$\,Hz and $U_{eg} = 0.8\,U_{gg}$ in this plot.
\textbf{(b)} Method for measuring $U_{ee}$. For strong Rabi frequencies and pulse area $\Omega T=2\pi$, the total population of doubly occupied sites has decreased due to inelastic losses (black solid line). This loss resonance is shifted with respect to the single-atom resonance (dashed blue line, pulse area $\Omega T=\pi$) by $(U_{ee}-U_{gg})/(2\hbar)$, with weak dependence on $U_{eg}$ (see inset). For illustrative purposes, $\Omega/(2\pi) = 1500$\,Hz and $U_{ee} = 2.5\,U_{gg}$ in this plot.
\textbf{(c)} Experimental determination of $U_{eg}$ with $\Omega T\approx \pi$. The ``shoulder'' near $\delta/(2\pi) \approx -300\,$Hz indicates the excitation of doubly occupied sites.
\textbf{(d)} Experimental determination of $U_{ee}$. The open symbols show $N_g$ for the reference measurement with $\Omega T\approx \pi$ locating the single-atom resonance. The closed ones correspond to $N_g+N_e$ for the loss measurement with $\Omega T \approx 2\pi$. The loss curve is almost centered on the single-atom resonance. 
A common fit to all data sets in \textbf{(c)} and \textbf{(d)} (solid and dashed lines) yields best fit parameters $U_{ee} = 0.97(23)\,U_{gg}$ and  $U_{eg} = 0.82(8)\,U_{gg}$ (see text). The quoted error bars are statistical $68\,$\% confidence intervals obtained by the bootstrap method. In all plots, zero detuning corresponds to the single-atom resonance.} 
\label{fig4}
\end{figure*}

We now turn to the determination of elastic interaction parameters $U_{eg}$ and $U_{ee}$. To this end, we perform spectroscopic experiments probing doubly-occupied sites. The method is illustrated in Fig. \ref{fig4}a-b, and the measurements shown in Fig. \ref{fig4}c-d. 

The determination of $U_{eg}$ is best performed in a perturbative limit, where the pulse area and the population of $\vert ee \rangle$ remain small (Fig. \ref{fig4}a). The time evolution of $\vert \Psi^{(2)} \rangle$ then reduces to that of a two-level system resonant for $\hbar\delta = U_{eg}-U_{gg}$. This resonance is well-resolved provided that the Rabi frequency is much smaller than $(U_{eg}-U_{gg})/\hbar$. 

In order to extract the interaction strength $U_{ee}$, one could in principle use a two-photon resonance directly linking $\vert gg \rangle$
and $\vert ee \rangle$. This requires a weak enough Rabi frequency $\Omega \ll \Delta$ and $\delta' \ll \Delta$, where $\hbar\Delta= U_{eg}-(U_{ee}+U_{gg})/2$ is an interaction shift and where $\delta'=\delta-(U_{ee}-U_{gg})/(2\hbar)$ is the two-photon detuning. Under these conditions, the intermediate state  $\vert eg \rangle$ can be adiabatically eliminated, and the dynamics reduces to that of an effective two-level system. The difference $(U_{ee}-U_{gg})/2$ can therefore be directly measured from the location of the two-photon resonance. Practically, this idealized experiment is difficult to perform for weak coupling due to the strong loss rate $\gamma_{ee}$, which gives a substantial width to the two-photon resonance. In order to circumvent this issue, we perform the experiment at a larger Rabi frequency, and make use of the losses by measuring $N_g+N_e$ after a clock pulse of area $\Omega T = 2\pi$ (Fig. \ref{fig4}b). The ``background signal'' from singly-occupied sites is minimized near resonance, whereas doubly-occupied sites show a pronounced feature due to $e-e$ losses located at $\hbar\delta \approx (U_{ee}-U_{gg})/2$. Even for large Rabi frequencies, we find that the loss spectral feature in the total signal is only weakly affected by the intermediate $\vert eg\rangle$ state (inset of Fig. \ref{fig4}b).

The experimental results are presented in Fig. \ref{fig4}c-d. Data is centered so that $\delta=0$ corresponds to the single-atom resonance. The measurement of $U_{eg}$ (Fig. \ref{fig4}c) is done with a weak Rabi frequency $\Omega^{\mathrm{weak}}/(2\pi) \approx 150\,$Hz and displays a ``shoulder'' near $\delta/(2\pi) \approx -300\,$Hz. This corresponds to the signal from doubly occupied sites. On the other hand, the measurement of $U_{ee}$ (Fig. \ref{fig4}d), performed at strong Rabi frequency $\Omega^{\mathrm{strong}}/(2\pi) \approx 1500\,$Hz, shows a loss peak almost coincident with $\delta/(2\pi) \approx 0\,$Hz, or equivalently $U_{ee} \approx U_{gg}$. To extract quantitative values, we fit the prediction of the model from Sec. \ref{sec:model} to the experimental spectra (solid and dashed lines in Fig. \ref{fig4}c-d). We fix the normalized populations $\mathcal{F}_{\overline{n}=1,2}$, the loss rates $\gamma_{ee}$, $\gamma_{eg}$ and the initial atom number $N_\mathrm{at}$ to their measured values and leave the Rabi frequencies $\Omega^{\mathrm{weak}}$, $\Omega^{\mathrm{strong}}$, the interaction energies $U_{ee}$, $U_{eg}$  and the repumping efficiency $\eta_{e}$ as free parameters. The prediction of the model has been further convolved with a Gaussian function to account for frequency jitter of the clock laser, the width $\sigma$ being left as an extra free parameter. We obtain $\Omega^{\mathrm{weak}}/(2\pi) = 145(13)$\,Hz, $\Omega^{\mathrm{strong}}/(2\pi) = 1470(70)$\,Hz and $\eta_{e} = 68(6)$\,\%, consistent with our calibrations. The width of the convolving Gaussian $\sigma = 100(40)$\,Hz, is consistent with the narrowest spectrum we could observe, as discussed in Sec. \ref{sec:singleatom}. Finally we extract $(U_{ee} - U_{gg})/h = -40(340)\,$Hz and $(U_{eg} - U_{gg})/h = -270(120)\,$Hz, in agreement with the qualitative discussion above. The error bars represent statistical $68\,$\% confidence intervals on the optimal values of $U_{eg}$ and $U_{ee}$, obtained by the bootstrap method.

\subsection{Determination of atomic parameters}

The ratio of the elastic interaction parameters $U_{eg}$ and $U_{ee}$ to $U_{gg}$ is directly proportional to the ratio of the respective scattering lengths. From our calibration of $U_{gg}$ (see appendix) we get 
\begin{eqnarray}
a_{eg}-a_{gg} & = & -19(11)\,a_0, \\
a_{ee}-a_{gg} & = & -3(25)\,a_0.
\end{eqnarray}
The error bars do not account for possible systematic errors (for instance, in determining $U_{gg}$). Combining our measurements and the value $a_{gg} = 105\,a_0$ from \cite{kitagawa2008a}, we obtain the scattering lengths $a_{eg} = 86(11)\,a_0$, $a_{ee} = 102(25)\,a_0$. We thus find all scattering lengths involving the clock states of ${}^{174}$Yb equal within 20\,\% \cite{hoffer2016}. The near-equality of the scattering lengths is somewhat surprising. These observations differ markedly from the fermionic $^{173}$Yb isotope, where the equivalent scattering lengths have been found very different from one another \cite{capellini2014a,scazza2014a}.

We also extract from the loss rate $\gamma_{ee}$ the two-body loss rate constant $\beta_{ee}$ that enters into the rate equation $\mathrm{d} \langle \hat{\Psi}_e^\dagger \hat{\Psi}_e \rangle /\mathrm{d}t = - \beta_{ee} \langle \hat{\Psi}_e^\dagger \hat{\Psi}_e^\dagger \hat{\Psi}_e \hat{\Psi}_e \rangle$ [the relation between the two is $\hbar\gamma_{ee}/U_{gg}=M\beta_{ee}/(4\pi\hbar a_{gg})$]. We find
\begin{eqnarray}
\beta_{ee} = 2.6(3)\times 10^{-11}\,\mathrm{cm}^3/\mathrm{s}.
\end{eqnarray}
This value is in line with comparable measurements in strontium or fermionic ytterbium \cite{scazza2014a,traverso2009a,ludlow2011a}. As noted in Sec. \ref{sec:losses}, we can only give an upper bound on the rate constant $\beta_{eg} \leq   10^{-15}\,\mathrm{cm}^3/\mathrm{s}$. Low $e-g$ inelastic loss rates were also observed for fermionic $^{173}$Yb \cite{scazza2014a} and $^{87}$Sr \cite{Bishof2011a}.

During the preparation of the manuscript, we learnt about similar experiments performed at LENS in Florence \cite{Franchi2017a}. Their results agree with ours within the uncertainties. 

\section{Conclusion}
\label{sec:conclusion}
In conclusion, we have performed spectroscopy of a bosonic ytterbium Mott insulator using an optical clock transition. The high spectral resolution allows us to be sensitive to the on-site number statistics. Singly-occupied sites display long-lived coherent oscillations between $g$ and $e$ whereas the dynamics of doubly-occupied sites is strongly affected by two-body elastic and inelastic interactions. We extract the inelastic loss constants from lifetime measurements, and the elastic interaction parameters from spectroscopy. The resulting intra- and inter-states scattering lengths $a_{gg}$, $a_{eg}$ and $a_{ee}$ turn out to be very close to one another.

The large inelastic loss rate for two atoms in $e$ at the same lattice site leads to sub-ms lifetimes. Such high rates are in line with measurements on other group-II atoms \cite{traverso2009a,ludlow2011a}, and are clearly a threat to experiments where such double occupancies can arise. Several solutions can be enforced to avoid this situation. If tunneling is irrelevant or detrimental (as in optical clocks), one can choose to work in a regime where only a Mott insulator with unit occupancy arises. This requires careful engineering of the auxiliary potential $V_\mathrm{T}$, but it seems within current experimental capability even for a large atom number of several $10^5$. In a regime where tunneling cannot be neglected, one could imagine using an ``interaction blockade'', where transitions to sites with double $e$ occupancy will always be off resonant and thereby inefficient. The near-equality of $a_{gg}$, $a_{eg}$ and $a_{ee}$ for ${}^{174}$Yb restricts this method to very low Rabi frequencies. Finally, the large inelastic rate suggests that the analog of the ``quantum Zeno'' suppression of losses \cite{syassen2008a,yan2013a,zhu2014a} could occur in our system, where strong losses confine the system to a lossless subspace for a suitable initial state and weak enough coupling. This provides a natural direction for future work in a true Hubbard regime where both interactions and tunneling play a role.

\textbf{Acknowledgments:}
We thank M. H\"ofer and S. F\"olling for discussions and for sharing their experimental results with us. We also thank G. Cappellini, J. Catani, L. Fallani and the LENS ytterbium team for stimulating discussions during the program ``From Static to Dynamical Gauge Fields with Ultracold Atoms'' (Galileo Galilei Institute, Florence). We are grateful to S. Nascimb{\`e}ne and J. Dalibard for their careful reading of the manuscript.

\bibliography{SpectroYb3DLattice_bib_varxiv}
%\nocite{*}
\bibliographystyle{unsrt}
\appendix*
\section{Collapse and revival experiment}
\label{app:cr}

\begin{figure}[!h]
\includegraphics[width=0.5\textwidth]{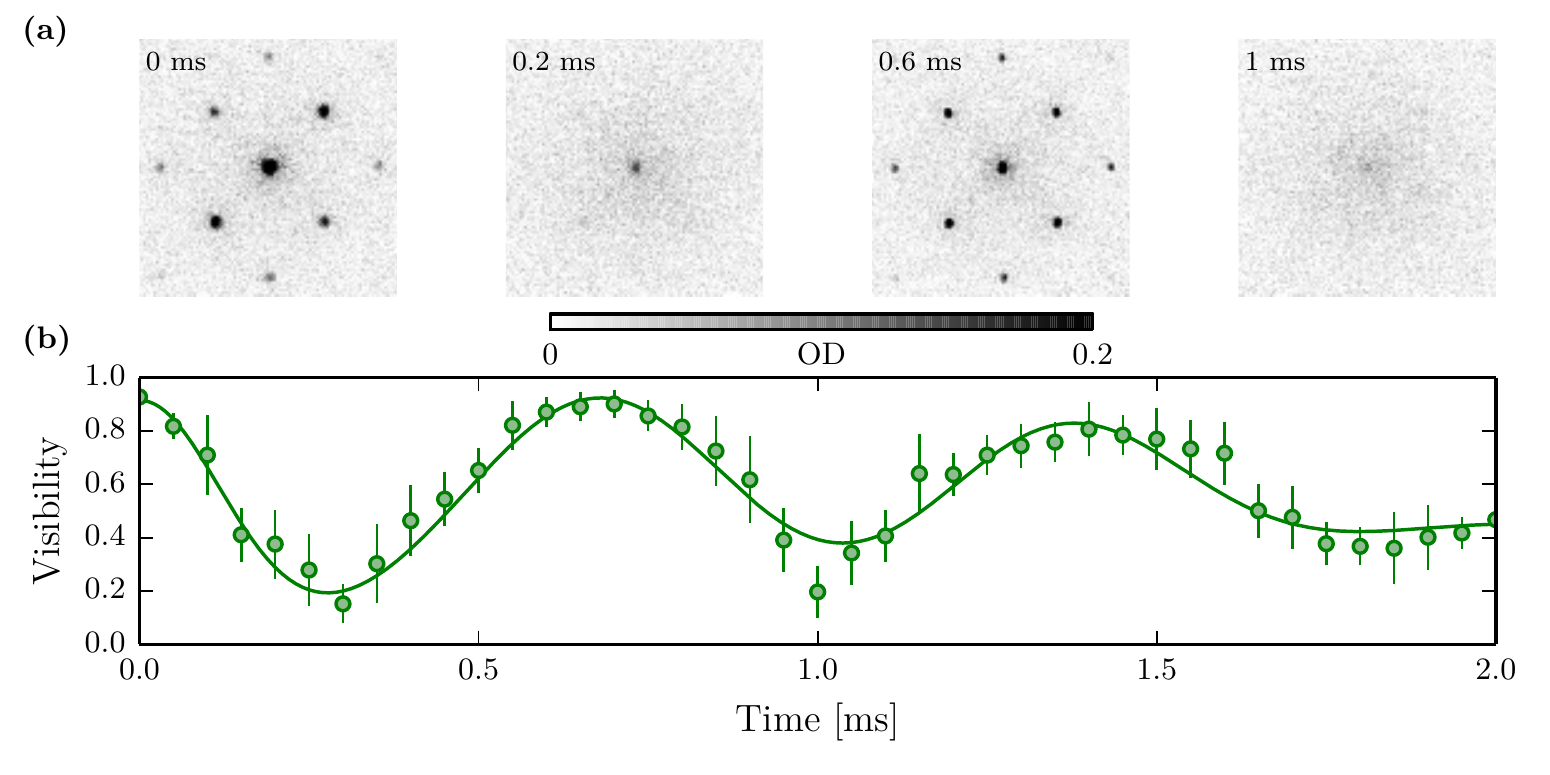}
\caption{\textbf{(a)} Experimental images of the interference pattern at different hold times after quenching the HL. \textbf{(b)} Visibility of the interference pattern undergoing collapse and revival dynamics. The period of the revivals is given by the on-site interaction energy $U_{gg}$. The solid line is a phenomenological fit using a sum of equidistant Gaussian functions giving $U_{gg}/h = 1475(25)\,$Hz.}
\label{figA1}
\end{figure}

We have performed a collapse and revival experiment following \cite{greiner2002b} to determine experimentally the interaction strength $U_{gg}$. We first prepare our system in a superfluid state by loading the atoms in a lattice with depths $V_{0,x} \approx V_{0,y} \approx 5\,E_\mathrm{R}$ and $V_{0,z}\approx 27 \,E_\mathrm{R}$. Rapidly quenching the HL depths to $V_{0,x} \approx V_{0,y} \approx 26\,E_\mathrm{R}$ suppresses tunneling in the $x-y$ plane, and triggers a coherent interaction-driven evolution where first-order spatial coherence periodically collapses and revives with period $h/U_{gg}$ (Fig. \ref{figA1}a-b). 
In a non-uniform system, the period of revivals is determined by $U_{gg}$ alone (which can be taken uniform over the lattice with an error smaller than 1\,\%), but the revival amplitude progressively decreases due to the inhomogeneous dephasing introduced by the auxiliary trap $V_\mathrm{T}$ \cite{greiner2002b}. To extract the period without detailed modeling of the system, we fit equidistant Gaussian functions to our experimental data (Fig. \ref{figA1}b). In this way we account phenomenologically for the inhomogeneous damping as in \cite{greiner2002b}.

\end{document}